\documentclass[aps,pra,twocolumn,superscriptaddress]{revtex4}

\usepackage[dvipdfmx]{graphicx}
\usepackage{bm}
\usepackage{amsmath,amssymb}
\usepackage{color}

\begin{document}

\title{
 Laser control of magnonic topological phases in antiferromagnets
}

\author{Kouki Nakata}
\affiliation{Advanced Science Research Center, Japan Atomic Energy Agency, Tokai 319-1195, Japan}
\author{Se Kwon Kim}
\affiliation{Department of Physics and Astronomy, University of Missouri, Columbia, Missouri 65211, USA}
\author{Shintaro Takayoshi}
\affiliation{Max Planck Institute for the Physics of Complex Systems, Dresden 01187, Germany}
\affiliation{Department of Quantum Matter Physics, University of Geneva, Geneva 1211, Switzerland}

\date{\today}

\begin{abstract}
We study the laser control of magnon topological phases induced by 
the Aharonov-Casher effect in insulating antiferromagnets (AFs).
Since the laser electric field can be considered as a time-periodic 
perturbation, we apply the Floquet theory and perform the inverse 
frequency expansion by focusing on the high frequency region. 
Using the obtained effective Floquet Hamiltonian,
we study nonequilibrium magnon dynamics away from the adiabatic limit
and its effect on topological phenomena.
We show that a linearly polarized laser can generate helical edge magnon states and induce the magnonic spin Nernst effect,
whereas a circularly polarized laser can generate chiral edge magnon states and induce the magnonic thermal Hall effect. 
In particular, in the latter, we find that the direction of the magnon chiral edge modes and the resulting thermal Hall effect can be controlled 
by the chirality of the circularly polarized laser through the change 
from the left-circular to the right-circular polarization.
Our results thus provide a handle to control and design magnon topological properties in the insulating AF.
\end{abstract}

\maketitle

\section{Introduction}
\label{sec:intro}

The utilization of the quantized spin wave, magnons, plays an increasingly important role in spintronics, 
spawning its subfield, magnon-spintronics a.k.a. magnonics 
\cite{MagnonSpintronics,ReviewMagnon}. The main subject in this 
field is the realization of rapid and efficient transmission of 
information through spins. 
For this purpose, antiferromagnets (AFs)~\cite{AFspintronicsReview2,AFspintronicsReview,AFreviewYT} 
have an advantage over ferromagnets (FMs)~\cite{demokritov,RShindou,RShindou2} in that the dynamics is much faster in the former since the former energy scale arising from microscopic and quantum-mechanical spin exchange interactions is much larger than the latter energy scale governed by the macroscopic magnetic 
dipole interaction. 

Another important viewpoint is the error-tolerance of 
communication, and topology is a useful tool to realize the states 
robust against impurities. 
In Ref.~\cite{KSJD}, 
a magnonic topological insulator (TI) is realized in the AF with 
the electric field gradient making use of the opposite magnon chirality~\cite{Kishine,Kishine2,AFnonabelian} associated with the N\'eel magnetic order. 
This gradient field behaves as the gauge potential for magnons through the 
Aharonov-Casher (AC) effect~\cite{casher} and forms the Landau level of magnons in the bulk.
In particular, magnons with the opposite magnon chirality carrying a magnetic dipole moment $\sigma  g\mu_{\rm{B}} {\bf e}_z$ with $\sigma = \pm 1$, 
where $\mu_{\rm{B}}$ is the Bohr magneton and $g$ is the $g$-factor of the constituent spins, 
propagate along the edge of the sample in the opposite direction and thus
the helical edge modes are realized in AFs, being in contrast to the 
chiral edge mode in FMs~\cite{KJD} characterized by the single  magnon chirality.
The spin transport properties in such AFs have a topological nature and 
cannot be disturbed by local perturbations. 
Thus the next task is to elucidate how to manipulate the topology in magnonic TIs.

\begin{figure}[t]
\centering
\includegraphics[width=0.45\textwidth]{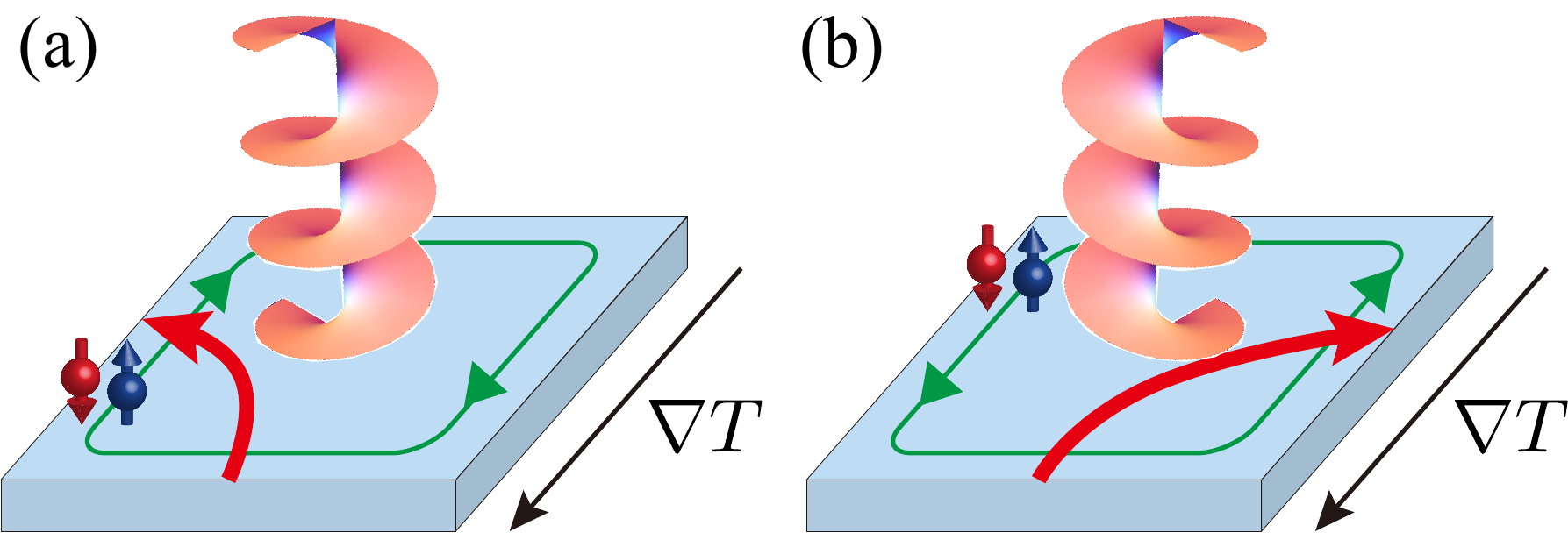}
\caption{
Schematic representation of magnon states in two-dimensional AFs 
subjected to a laser. A circularly polarized laser (pink-colored spiral) 
generates a pair of chiral edge magnon states and induces magnonic thermal Hall 
effect where up and down magnons ($\sigma =\pm 1$ represented by blue 
and red balls with arrows, respectively) with opposite magnetic dipole 
moments $\sigma  g\mu_{\rm{B}}  {\bf e}_z$ propagate along the edge of 
a finite size sample in the same direction. The direction of the magnon chiral 
edge modes (green lines with arrows) and the resulting thermal Hall current 
(red arrows) can be controlled by changing the chirality of the 
circularly polarized laser between the (a) right-circular $\eta = -1 $ and (b) left-circular $\eta = 1 $ polarization.
}
\label{fig:Setup}
\end{figure}

A conventional way to change the physical state is tuning the control 
parameters of the system, e.g., temperature, pressure, and a static 
electromagnetic field. However a remarkable advance in the field of 
quantum optics offers a novel method to the manipulation of the state; the application of laser. A number of studies have been conducted both 
theoretically and experimentally on the laser-induced or -controlled states such as the photoinduced metal-insulator transition~\cite{Iwai2003PRL,Ishikawa2014NatComm}, Floquet topological phases~\cite{FloquetOkaSan,FloquetOkaSan2,FloquetOkaSan4,FloquetJK,FloquetJK2}, and laser-induced magnetic states~\cite{Kimel2005Nature,Kirilyuk2010RMP,FloquetST,FloquetST2,FloquetST3}.
In particular, for magnetic systems, the typical energy scale 
is on the order of meV, which corresponds to the terahertz frequency. Thus spin 
manipulation is performed in the picosecond time interval and it is much 
faster than the time scale of conventional spintronics. Therefore 
establishing the way to control magnonic TIs~\cite{KSJD} is an 
essential ingredient for the ultrafast topological magnonics.

In this paper, we consider the application of a laser to the insulating AF 
on the square lattice with an easy-axis magnetic anisotropy which hosts 
two kinds of gapped magnons with the same parabolic dispersion and the opposite 
magnetic dipole moment~\cite{KSJD}. At low temperature~\cite{magnon10mK}, 
interaction effects such as magnon-magnon and magnon-phonon interactions 
become negligibly small~\cite{magnonWF,adachiphonon,Tmagnonphonon}. 
We treat the effect of the laser as a time-periodic electric field, which is incorporated 
into the Hamiltonian as the time-periodic AC potential. This system can 
be analyzed by the Floquet theory and in the high frequency regime, 
we can obtain the effective Floquet Hamiltonian~\cite{FloquetReview,FloquetOkaSan4} by the high frequency expansion.
We find that the linearly polarized laser with nonzero time-averaged field 
induces helical magnon edge states, while the circularly polarized laser 
induces a pair of chiral magnon edge states (Fig.~\ref{fig:Setup}). 
The laser response depending on the direction of the magnetic dipole moment 
for each magnon plays an important role. 
In another perspective, our study corresponds to the further extension of 
Ref.~\cite{KSJD} into the nonequilibrium regime away from the adiabatic 
limit. The resulting difference in the thermomagnetic properties 
\cite{katsura,Matsumoto,Matsumoto2,SMreviewMagnon,ohnuma,MagnonNernstAF,MagnonNernstAF2} of Hall transport is also discussed. 

This paper is organized as follows.
In Sec.~\ref{sec:magTI}, we quickly review the magnonic TI in the static 
situation. In Sec. \ref{sec:system}, considering three types of lasers, we derive each effective Floquet Hamiltonian and find the difference in the magnon motion focusing on the $\sigma$-dependence.
The resulting difference in thermomagnetic properties of Hall transport is discussed in Sec. \ref{sec:WF}.
In Sec. \ref{sec:Exp}, we provide a theoretical insight into experiments.
Finally, we remark on several issues in Sec. \ref{sec:discussion} and summarize in Sec. \ref{sec:summary}.
Technical details are described in the Appendix.

\section{Magnonic TI}
\label{sec:magTI}

Before considering the laser application, 
we quickly review the magnonic TI realized in the insulating AFs 
with the static electric field gradient. 
For the details, see Ref.~\cite{KSJD}.
The contents in this section are the basis for our study on the case of 
time-dependent electric field instead of the static field, which is 
discussed in the following sections.

It has been established that the spin-wave theory~\cite{BlochMagnon,KittelFMR} and its quantized version, the magnon picture \cite{HP}, well describes the thermomagnetic properties such as magnetization and specific heat in the AFs~\cite{AndersonAF,RKuboAF} as well as FMs. 
We consider the insulating AF on a two-dimensional square 
lattice residing in the $xy$ plane with magnetic anisotropy that prefers 
the $S^{z}$ axis. The ground state of this system has the N\'eel order 
along the $z$ direction and the low energy excitation 
structure is dictated in terms of as electrically neutral bosonic 
quasiparticles after the Bogoliubov transformation. 
Here there exist two kinds of bosons carrying a magnetic dipole moment $\sigma 
g\mu_{\rm{B}} {\bf e}_z$ with $\sigma\equiv \delta S^{z} = 1 (-1)$, 
which are respectively identified with up (down) magnons.
Due to the presence of easy-axis magnetic anisotropy, 
the insulating AFs have gapped and parabolic dispersion under the long 
wave-length approximation, and the dynamics can be described by using
the decoupled two magnon modes ($\sigma = \pm 1$) at temperature lower 
than the magnon gap. In the low-energy regime, such an antiferromagnetic 
magnonic system effectively preserves the time-reversal symmetry (TRS).

In Ref.~\cite{KSJD}, under the assumption that the total spin along the 
$z$ axis $\sum_{j}S_{j}^{z}$ is conserved and remains a good quantum number, 
we have proposed a magnonic analog of the quantum spin Hall effect 
characterized by helical edge states and thus established a bosonic counterpart of TIs~\cite{TIreview,TIreview2}, namely the magnonic TIs in insulating AFs using the above-mentioned picture for the clean systems and following the work by Aharonov and Casher~\cite{casher,footnote1}. 
The proposal is built upon the fact that an electric field couples to the magnetic dipole moment $\sigma g \mu _{\rm{B}}{\mathbf{e}}_z$ through the AC effect \cite{casher,Mignani,BalaAC,magnon2,KKPD,KPD,ACatom,AC_Vignale,AC_Vignale2,ACspinwave}, which is analogous to the Aharonov-Bohm (AB) effect \cite{bohm,LossPersistent,LossPersistent2} of electrically charged particles in magnetic fields.
Each magnon ($\sigma=\pm 1$) of the insulating AF subjected to a dc 
electric field with $ {\mathcal{E}} $
a constant gradient ${\mathbf{E}}({\mathbf{r}})= {\mathcal{E}}(-x, 0,0)$ as a function of the position ${\mathbf{r}}=(x,y,0)$  
experiences the ``electric'' vector potential~\cite{magnon2,KJD,KSJD}  ${\mathbf{A}}_{\rm{m}}({\mathbf{r}}) = {\mathbf{E}}({\mathbf{r}})\times {\mathbf{e_z}}/c= {{\mathcal{E}}}/{c}(0, x,0)$.
The decoupled Hamiltonian for each magnon ($\sigma\pm 1$) is represented as~\cite{KSJD}
\begin{align}
 {\cal{H}}_{\sigma } = \frac{1}{2m} \Big(\hat{{\mathbf{p}}} + \sigma \frac{g \mu _{\rm{B}}}{c}{\mathbf{A}}_{\rm{m}} \Big)^2
+ \Delta ,
\label{HamiltonianAC} 
\end{align}
where $\hat{{\mathbf{p}}} =(p_x, p_y, 0)$ is the momentum operator, $m$ 
is the effective mass of magnons, $\Delta$ is the magnon gap induced by 
easy-axis spin anisotropy. See Ref.~\cite{KSJD} for the specific expression of $m$ and $\Delta$ in terms of the spin language. 
The total Hamiltonian of the system is given by $\sum_{\sigma}{\cal{H}}_{\sigma}$, which respects the TRS effectively in the low-energy regime.
The $\sigma$-dependence stems from the opposite magnetic dipole moments $ \sigma  g \mu _{\rm{B}}{\mathbf{e}}_z$ of up and down magnons associated with the N\'eel order in insulating AFs.
This $\sigma$-dependence is the key ingredient~\cite{Kishine,Kishine2,AFnonabelian} for qualitatively new phenomena in AFs 
which are not found in FMs~\cite{KJD} such as 
the violation of the magnonic Wiedemann-Franz (WF) law for Hall transport~\cite{KSJD,KJD} and the generation of helical edge magnon states.
Experiencing the AC vector potential ${\mathbf{A}}_{\rm{m}}$ with ${\mathbf{\nabla }} \times {\mathbf{A}}_{\rm{m}} = ({{\mathcal{E}}}/{c}) {\mathbf{e}}_z$, magnons of opposite spins form the same Landau levels~\cite{KSJD} and performs cyclotron motions with the same frequency $\omega_c=(g \mu _{\rm{B}}/{m c^2}){\mathcal{E}}$ and with the same electric length $ l_{\rm{{\mathcal{E}}}} \equiv  \sqrt{{\hbar c^2}/{g \mu 
_{\rm{B}}{\mathcal{E}}}} $, but in the opposite direction, leading to the helical edge magnon state~\cite{footnote2}.
Note that the TRS as well as the total spin 
conservation along the $z$ axis protect the topological phase and 
helical edge states against nonmagnetic impurities. The key ingredient for the generation of topological edge states is the cyclotron motion in the bulk of the system where up and down magnons are decoupled.

It has been established theoretically that magnonic TIs are realized in 
the gradient dc electric field. However, whether those states remain intact or not in nonequilibrium, in other words, whether those topological properties are robust against time-dependent perturbation, is still an open issue and the scope of this paper.

\section{Magnon motion in laser}
\label{sec:system}

Applying the laser with a frequency $\Omega$ (a period $T = 2\pi/\Omega$) to the insulating AF described by Eq.~\eqref{HamiltonianAC},
magnons of opposite spins ($\sigma = \pm 1$) subjected to a periodic electric field ${\mathbf{E}}(t) = {\mathbf{E}}(t+T)$ 
acquire the AC vector potential~\cite{casher,magnon2,KJD},
\begin{align}
{\mathbf{A}}_{\rm{m}}(t) = {\mathbf{A}}_{\rm{m}}(t+T),
\label{FloquetHamiltonianAC2} 
\end{align}
and the Hamiltonian Eq.~\eqref{HamiltonianAC} also becomes time-periodic 
${\cal{H}}_{\sigma }(t)  = {\cal{H}}_{\sigma }(t+T) $. The decoupled 
Hamiltonian for each magnon ($\sigma=\pm 1$) is 
specifically written as
\begin{align}
 {\cal{H}}_{\sigma }(t)  =    \frac{1}{2m} \Big(\hat{{\mathbf{p}}} +  \sigma  \frac{g \mu _{\rm{B}}}{c}{\mathbf{A}}_{\rm{m}}(t) \Big)^2
 + \Delta .
\label{FloquetHamiltonianAC} 
\end{align}
and the total Hamiltonian of the system is given as $\sum_{\sigma}{\cal{H}}_{\sigma }(t)$.
Assuming that the system experiences a time-evolution away from the 
adiabatic limit with the application of high frequency laser $\Omega \gg 
 \omega_{\rm{c}}$, the magnon motion is described by the effective 
 Floquet Hamiltonian~\cite{FloquetReview} of 
Eq.~\eqref{FloquetHamiltonianAC}, which is represented as 
$ {\cal{H}}_{\rm{eff}}= \sum_{n =0}^{\infty } {\cal{H}}_{\rm{eff}}^{(n)}$, 
where the effects of the laser are taken into account perturbatively 
($1/\Omega^n$) via each component ${\cal{H}}_{\rm{eff}}^{(n)}$ of the 
high frequency expansion (see Appendix \ref{sec:review3} for details).

We remark that though the magnetic field is accompanied by the laser electric field, 
it does not couple to the orbital motion of magnons, i.e., the linear momentum of magnons 
but enters the magnon energy directly through the Zeeman coupling without affecting orbital motion. 
After time-averaging, the effect of the time-varying magnetic field on the magnon energy can be captured by renormalizing the energy gap of magnons,
and up and down magnons are still degenerate due to the easy-axis spin anisotropy and the resultant magnon energy gap~\cite{footnote4}.
On the contrary, an electric field affects the orbital motion of magnons via AC effects and thereby can induce the finite Berry phases for magnons as studied in Refs.~\cite{KJD,KSJD}. 
Thus we study the effects of the coupling of an ac electric field of laser to orbital motion of magnons on the magnon bands, looking for possible laser-induced topological phases of magnons.
In this paper, our consideration is restricted to the magnon dynamics in 
the high frequency region $\Omega \gg  \omega_c$, where 
the Floquet Hamiltonian \cite{FloquetReview} of Eq.~\eqref{FloquetHamiltonianAC} can be analyzed through the high-frequency expansion.

\subsection{Linearly polarized laser}
\label{subsec:system1}

First let us consider the case of linearly polarized laser providing the 
electric field ${\mathbf{E}}(t)= {\mathcal{E}}(-x {\rm{cos}}(\Omega t), 
0,0)$. This gives rise to the periodic AC vector potential 
\begin{align}
{\mathbf{A}}_{\rm{m}}(t) = \frac{\cal{E}}{c}(0, x {\rm{cos}}(\Omega t), 0),
\label{A1} 
\end{align}
which is time-reversal invariant ${\mathbf{A}}_{\rm{m}}(t) = {\mathbf{A}}_{\rm{m}}(-t) $,
and the time averaged value is zero $ \bar{{\mathbf{A}}}_{\rm{m}}(t) = 0  $.
After the high frequency expansion up to ${\cal{O}}(1/\Omega ^2)$, 
we obtain the effective Floquet Hamiltonian as 
${\cal{H}}_{\rm{eff}}={\cal{H}}_{\rm{eff}}^{(0)} 
+{\cal{H}}_{\rm{eff}}^{(2)}$ (see Appendix \ref{sec:review3} for details), where
\begin{align}
\begin{split}
{\cal{H}}_{\rm{eff}}^{(0)}=& \frac{1}{2m}\Big[p_x^2 + p_y^2 + \frac{1}{2} \Big(\frac{g \mu_{\rm{B}}}{c} \Big)^2  
 \Big(\frac{\cal{E}}{c} \Big)^2 x^2 \Big]+ \Delta,\\
{\cal{H}}_{\rm{eff}}^{(2)} =& \Big(\frac{1}{2m}\Big)^3 \frac{1}{\Omega^2}\Big[2 \Big(\frac{g \mu_{\rm{B}}}{c} \Big)^2   \Big(\frac{\cal{E}}{c} \Big)^2 p_y^2\\
&\qquad\qquad+\frac{1}{8} \Big(\frac{g \mu_{\rm{B}}}{c} \Big)^4   \Big(\frac{\cal{E}}{c} \Big)^4 x^2 \Big].
\end{split}
\label{Heff_linearI}
\end{align}
The cancellation of the $ {\cal{H}}_{\rm{eff}}^{(1)}$ term reflects the 
time-reversal invariance ${\mathbf{A}}_{\rm{m}}(t) = 
{\mathbf{A}}_{\rm{m}}(-t)$. 
From Eq.~\eqref{Heff_linearI}, we find that 
the effective magnon mass is renormalized as for the motion along the 
$y$ direction and the confinement by the harmonic potential happens 
along the $x$ direction, and both effects are irrelevant with topological properties of magnons.
Since there are no terms coupling momentum and spatial coordinates such 
as $p_y x$ and $p_x y$ which play the role of the Lorentz 
force~\cite{footnote5} for magnons, the Landau energy level 
\cite{KJD,KSJD} is not formed and magnons do not perform the cyclotron 
motion, leading to the absence of any magnon edge states in the high 
frequency region.
Therefore the linearly polarized laser Eq.~\eqref{A1} does not bring any topological properties to magnon transport in the insulating AFs;
the absence of any edge magnon states and the topologically trivial bulk without any magnon Hall effects~\cite{magnon2,KJD,KSJD}.

\subsection{Linearly polarized laser with nonzero time-averaged field}
\label{subsec:system2}

In this section we consider another type of linearly polarized laser ${\mathbf{E}}(t)= {\mathcal{E}}(-x {\rm{cos}}^2(\Omega t), 0,0)$.
The resulting periodic AC vector potential is given by
\begin{align}
{\mathbf{A}}_{\rm{m}}(t) = \frac{\cal{E}}{c}(0, x {\rm{cos}}^2(\Omega t), 0),
\label{A2} 
\end{align}
being time-reversal invariant ${\mathbf{A}}_{\rm{m}}(t) = {\mathbf{A}}_{\rm{m}}(-t)$,
whereas in contrast to the case of Sec.~\ref{subsec:system1} the time averaged value becomes nonzero 
$   \bar{A} _{\rm{m}}^y (t) =  ({\cal{E}}/2c) x $.
After the high frequency expansion ($\Omega \gg  \omega_{\rm{c}}  $), an 
effective Floquet Hamiltonian ${\cal{H}}_{\rm{eff}}={\cal{H}}_{\rm{eff}}^{(0)} +{\cal{H}}_{\rm{eff}}^{(2)}$
up to ${\cal{O}}(1/\Omega ^2)$ is derived as 
(see Appendix \ref{sec:review3} for details),
\begin{align}
 {\cal{H}}_{\rm{eff}} = \frac{1}{2m}\Big[ p_x^2 + (1+t_0)\Big(p_y + \sigma \frac{g \mu_{\rm{B}}}{c} \frac{{\cal{E}}_{\rm{eff}}}{c}x \Big)^2 \Big] + \Delta,
 \label{Heff2total}
\end{align}
where
\begin{align}
 {\cal{E}}_{\rm{eff}} = \frac{1+2 t_0}{1+t_0} \frac{{\cal{E}}}{2} 
\label{Eeff2}
\end{align}
is the effective electric field gradient and 
$t_0=(1/32)({\omega _c}/{\Omega})^2  \propto  1/{\Omega^2}$.
Here the harmonic potential term is dropped since it is irrelevant to 
topological properties of magnons.
Due to the emergence of the effective electric field gradient ${\cal{E}}_{\rm{eff}}$,
magnons of opposite spins ($\sigma =\pm 1$) form the same Landau energy level and perform the cyclotron motion~\cite{KJD,KSJD} with the same frequency 
and the same electric length but in the opposite direction depending on 
$\sigma$ as is seen from Eq.~\eqref{Heff2total}, 
which leads to helical edge magnon states.

Note that the total spin conservation and the TRS
still holds in the present setup. There symmetries protect the topological phase and helical edge states against nonmagnetic impurities~\cite{footnote3}.

\subsection{Circularly polarized laser}
\label{subsec:system3}

Next we move on to the case of circularly polarized laser 
\cite{FloquetOkaSan,FloquetDiracFermion,FloquetDiracFermion2}. The laser 
electric field is ${\mathbf{E}}(t)= {\mathcal{E}}(-x 
{\rm{cos}}(\Omega t), \eta  x {\rm{sin}}(\Omega t),0)$, where $\eta = 1 
(-1)$ is the index to represent the left (right) circular polarization.
Then the periodic AC vector potential becomes 
\begin{align}
{\mathbf{A}}_{\rm{m}}(t) = \frac{\cal{E}}{c}(\eta  x {\rm{sin}}(\Omega t), x {\rm{cos}}(\Omega t), 0),
\label{A3} 
\end{align}
where the time averaged value vanishes $\bar{{\mathbf{A}}}_{\rm{m}}(t)=0$ 
as is the case of Sec.~\ref{subsec:system1}, 
though the time-reversal invariance is violated by the circularly 
polarized laser ${\mathbf{E}}(t) 
\neq {\mathbf{E}}(-t)$ and ${\mathbf{A}}_{\rm{m}}(t) \neq {\mathbf{A}}_{\rm{m}}(-t) $~\cite{FloquetST,FloquetST2,FloquetST3} 
in contrast to the case of Secs. \ref{subsec:system1} and \ref{subsec:system2}.
Again we obtain an effective Floquet Hamiltonian up to ${\cal{O}}(1/\Omega )$ 
using the high frequency expansion ($\Omega \gg  \omega_{\rm{c}}$) as 
${\cal{H}}_{\rm{eff}}={\cal{H}}_{\rm{eff}}^{(0)} 
+{\cal{H}}_{\rm{eff}}^{(1)}$ (see Appendix \ref{sec:review3} for details), where
\begin{subequations}
\begin{align}
{\cal{H}}_{\rm{eff}}^{(0)} =& \frac{1}{2m}\Big[p_x^2 + p_y^2 +  \Big(\frac{g \mu_{\rm{B}}}{c} \Big)^2   \Big(\frac{\cal{E}}{c} \Big)^2 x^2 \Big]+ \Delta ,
\label{H_Effm30}\\
{\cal{H}}_{\rm{eff}}^{(1)} =& -  \eta  \frac{\omega_{\rm{c}}}{2 \Omega} \omega_{\rm{c}}  p_y x.
\label{H3eff}
\end{align}
\end{subequations}
After dropping the harmonic potential term for readability due to its irrelevance with topological properties of magnons,
the Hamiltonian is recast into 
\begin{subequations}
\begin{align}
 {\cal{H}}_{\rm{eff}} =& \frac{1}{2m}\Big[ p_x^2 + \Big(p_y - \eta  \frac{g \mu_{\rm{B}}}{c} \frac{{\cal{E}}_{\rm{eff}}}{c}x \Big)^2 \Big]  
+ \Delta ,  
  \label{Heff3E}   \\
 {\cal{E}}_{\rm{eff}} =& \frac{\omega_c}{2 \Omega } {{\cal{E}}} \propto  \frac{1}{\Omega}.
\label{Eeff3}  
\end{align}
\end{subequations}
Contrary to Secs. \ref{subsec:system1} and \ref{subsec:system2},
since the circularly polarized laser ${\mathbf{E}}(t) \neq {\mathbf{E}}(-t)$
breaks the TRS (${\mathbf{A}}_{\rm{m}}(t) \neq 
{\mathbf{A}}_{\rm{m}}(-t)$), 
the leading correction does not vanish $ {\cal{H}}_{\rm{eff}}^{(1)}\not=0$
and brings the term proportional to $p_y x$ [Eq.~\eqref{H3eff}] that works as a Lorentz force for magnons \cite{magnon2,KJD,KSJD}, namely, 
the laser-induced Lorentz force proportional to $1/\Omega$.
Note that the force does not depend on the index $\sigma =\pm 1$ for up 
and down magnons, while it depends on the index $ \eta = 1 (-1)$ for the 
left (right) circular polarization of the laser. Therefore both up and 
down magnons perform the cyclotron motion in the same direction 
depending on the sign $\eta$, differently from the case of the linearly polarized laser in Sec.~\ref{subsec:system2}. 
In other words, the direction of the cyclotron motion can be controlled by tuning the chirality of the circularly polarized laser.

To conclude, in the circularly polarized laser magnons acquire the laser-induced effective electric field gradient ${\cal{E}}_{\rm{eff}}\propto  1/\Omega$. 
Thereby forming the same Landau energy level, 
magnons of opposite spins ($\sigma = \pm 1$) perform the cyclotron motion along the same direction, leading to the chiral edge magnon states. 
The direction of the cyclotron motion and that of the resulting chiral 
edge magnons can be controlled by changing the chirality of the 
circularly polarized laser between the left-circular or the 
right-circular polarization ($\eta = \pm 1$) as can be seen from 
Eqs.~\eqref{Heff3E} and \eqref{Eeff3}, which are the main results of 
this paper. The schematic figure of the present setup and induced chiral 
magnon edge modes is shown in Fig.~\ref{fig:Setup}. 

Since the effective gradient electric field [Eq.~\eqref{Eeff3}] induced by circularly polarized laser vanishes 
in the high frequency limit $\Omega \to \infty$,
it is interpreted as an `emergent' field having an intrinsically
nonequilibrium nature away from the adiabatic limit $\Omega \to 0$.

\begin{table*}
\caption{
\label{tab:table1}
High frequency laser-induced magnon motion and the thermomagnetic properties of magnon Hall transport
in the insulating AF [Eq.~\eqref{FloquetHamiltonianAC}]. 
}
\begin{ruledtabular}
\begin{tabular}{l|ccc}
Laser of high frequency $ \Omega \gg   \omega_c$
&  Linearly polarized laser:
&  Linearly polarized laser: 
&  Circularly polarized laser:        \\  
& Sec. \ref{subsec:system1} 
& Sec. \ref{subsec:system2}
& Sec. \ref{subsec:system3}       \\  
&    
$ {\mathbf{A}}_{\rm{m}}(t) = {\mathbf{A}}_{\rm{m}}(-t) $  &  
$ {\mathbf{A}}_{\rm{m}}(t) = {\mathbf{A}}_{\rm{m}}(-t) $  &  
$ {\mathbf{A}}_{\rm{m}}(t) \not= {\mathbf{A}}_{\rm{m}}(-t) $     \\ 
&    
$ \bar{{\mathbf{A}}} _{\rm{m}}(t) =  0$  &  
$ \mid   \bar{{\mathbf{A}}} _{\rm{m}}(t)   \mid  =  ({\cal{E}}/2c)x $  &  
$  \bar{{\mathbf{A}}} _{\rm{m}}(t) =  0$    \\  \hline  
 ${\cal{H}}_{\rm{eff}}$   & 
$\sigma$-independent    &  
$\sigma$-dependent   &
$\sigma$-independent    \\
& 
   \    &  
     \   &
$\eta$-dependent   \\
 $ {\cal{E}}_{\rm{eff}}$ &    
$ 0$ &  
$ [(1+2 t_0)/(1+t_0)] ({\cal{E}}/2)$   &  
$  (\omega_c/2\Omega){\cal{E}}$  \\ 
 Topological properties  & 
$-$    &  
 \checkmark  &
   \checkmark  \\
 Cyclotron motion of each mode & 
$-$    &  
 In the opposite direction   &
 In the same direction  \\
 Edge states  & 
$-$    &  
Helical edge states   &
Chiral edge states   \\  
Magnon thermal Hall effect  & 
  $-$   &  
 $-$  &
    \checkmark   \\    
 Magnon spin Nernst effect   & 
  $-$  &  
  \checkmark  &
  $-$  \\    
 Magnonic WF law \cite{KJD} &  
No Hall effects      &  
$-$       &
 \checkmark 
\end{tabular}
\end{ruledtabular}
\end{table*}

\subsection{Laser-driven magnon and symmetry}
\label{subsec:system4}

In the high frequency regime $ \Omega \gg  \omega_c $
the linearly polarized laser [Eq. (\ref{A2})] can induce the helical 
edge magnon states, while the circularly polarized laser [Eq. (\ref{A3})] can generate the chiral edge magnon states
whose direction can be controlled by changing the chirality of the laser, i.e., depending on the index $\eta = 1 (-1)$ for the left (right) circular [Eq. (\ref{Heff3E})].
Those insulating AFs in laser become topologically nontrivial.
Thus depending on the form of the laser, e.g.,  polarized linearly or circularly,
the details of the edges states (i.e., chiral or helical) in the topological AFs vary from system to system.
This indicates that by tuning the laser, we can control and design topological properties of antiferromagnetic magnonic systems.

We remark that since
${\mathbf{A}}_{\rm{m}}(t) = {\mathbf{A}}_{\rm{m}}(-t)$ for the linearly polarized laser,
those systems described by the Hamiltonian ${\cal{H}}_{\sigma }(t)$ [Eq. (\ref{FloquetHamiltonianAC})]
and the ones by the effective Floquet Hamiltonian ${\cal{H}}_{\rm{eff}}={\cal{H}}_{\rm{eff}}^{(0)} +{\cal{H}}_{\rm{eff}}^{(2)}$
possess the TRS. 
The TRS of the system can be seen by the transformation 
$\mathbf{p}\to -\mathbf{p}$ and $\sigma\to -\sigma$. 
On the other hand, since ${\mathbf{A}}_{\rm{m}}(t) \neq {\mathbf{A}}_{\rm{m}}(-t)$ 
for the circularly polarized laser,
the TRS is broken in the system described by the Hamiltonian ${\cal{H}}_{\sigma }(t)$ and
in the one by the effective Floquet Hamiltonian ${\cal{H}}_{\rm{eff}}={\cal{H}}_{\rm{eff}}^{(0)}  +{\cal{H}}_{\rm{eff}}^{(1)}$ [Eq.~\eqref{Heff3E}].
This TRS breaking stems from the chirality dependence $\eta = \pm 1 $.
Those results can be interpreted from the general properties of the Floquet 
formalism~\cite{FloquetReview}. 
When the Hamiltonian $ {\cal{H}} (t)  = \sum_{m \in{\mathbb{Z}}} H_{m} {\rm{e}}^{i m \Omega t}$
possesses the TRS $\mathcal{H} (t)=\mathcal{H} (-t)$, 
$[H_m, H_{-m}]$ becomes zero due to $H_m=H_{-m}$. 
Thus the $1/\Omega$ order term of the effective Floquet Hamiltonian 
$  {\cal{H}}_{\rm{eff}}^{(1)} =  ({\hbar \Omega })^{-1} \sum_{m=1}^{\infty } {[H_m, H_{-m}]}/{m}$
vanishes .
In contrast, when the TRS is broken, the ${\cal{H}}_{\rm{eff}}^{(1)}$ term [Eq. (\ref{H3eff})] 
can be nonzero. 


\section{Hall transport with the application of laser}
\label{sec:WF}

In this section, we discuss the laser-induced thermomagnetic properties of Hall transport in the topological AFs. 
Within the linear response regime, the spin and heat Hall current 
densities for each mode ($\sigma =\pm 1$) in the topological AFs 
subjected to an effective  magnetic field gradient (i.e., a gradient of nonequilibrium magnon chemical potential \cite{YacobyChemical}) 
and a temperature gradient are described by the Onsager matrix of Eq.~(31) in Ref.~\cite{KSJD}. 
Within the almost flat band approximation~\cite{KJD,KSJD,RSdisorder,KevinHallEffect}, the Onsager coefficients become characterized by the topological invariant 
(i.e., Chern integer) that edge states bring about. 

Since the linearly polarized laser [Eq.~\eqref{A2}] can induce helical edge magnon states, the total Chern number vanishes, while the ${\mathbb{Z}}_{2}$ topological invariant becomes nonzero~\cite{KSJD}. 
Therefore the diagonal elements of the Onsager matrix vanishes,
whereas the off-diagonal elements becomes nonzero.
This leads to the generation of the magnonic spin Nernst effect, 
while the vanishment of the magnonic thermal Hall effect. 
The vanishment of the magnonic spin Hall conductance $ G_{\rm{AF}}^{yx } =0$ and the thermal Hall conductance $ K_{\rm{AF}}^{yx } =0$ in the AFs indicate that the thermomagnetic ratio $ K_{\rm{AF}}^{yx }/ G_{\rm{AF}}^{yx } $ becomes ill-defined due to $G_{\rm{AF}}^{yx } = 0$ and that the WF law \cite{WFgermany,magnonWF,KJD} characterized by the liner-in-$T$ behavior at low temperature becomes violated due to $K_{\rm{AF}}^{yx } = 0$~\cite{KSJD}.

On the other hand, since the circularly polarized laser [Eq.~\eqref{A3}] can induce chiral edge magnon states \cite{KJD}
the ${\mathbb{Z}}_{2}$ topological invariant \cite{KSJD} vanishes, while the total Chern number becomes nonzero.
Therefore the diagonal elements of the Onsager matrix~\cite{KSJD}
becomes nonzero,
whereas the off-diagonal elements vanishes.
This leads to the generation of the magnonic thermal Hall effect, 
while the vanishment of the magnonic spin Nernst effect.
The direction of the thermal Hall current can be controlled 
by switching the chirality of the laser between the left-circular and 
right-circular polarization ($\eta = \pm 1$) [Eq.~\eqref{Heff3E}] as 
shown in Fig.~\ref{fig:Setup}.
The thermomagnetic ratio satisfies the magnonic WF law \cite{KJD} at low temperature \cite{magnon10mK},
\begin{align}
\frac{K_{\rm{AF}}^{yx }}{G_{\rm{AF}}^{yx }}  \stackrel{\rightarrow }{=}   \Big(\frac{k_{\rm{B}}}{g \mu _{\rm{B}}}\Big)^2    T,
\label{Eq:WFmagnon}
\end{align}
as the topological FM \cite{KJD} does satisfy.
Note that the thermal Hall effect of magnons has been observed in Ref.~\cite{onose}
and the measurement of a magnonic spin conductance has been reported in Ref.~\cite{MagnonG} where the gradient of a nonequilibrium magnonic spin chemical potential \cite{SilsbeeMagnetization,Basso,Basso2,Basso3,MagnonChemicalWees,YacobyChemical} plays the role of an effective magnetic field gradient. 
Thereby we expect that the magnonic WF law \cite{KSJD,KJD,magnonWF} can be experimentally confirmed \cite{ACspinwave,onose,ISHE1,spinwave,uchidainsulator,SekiAF,MagnonNernstExp,MagnonHallEffectWees,MagnonG,Snell_Exp,Snell2magnon,Camera2,demokritovReport}.

To conclude, depending on the form of laser such as linear or circular polarization,
the thermomagnetic properties of Hall transport (e.g., the magnonic WF law) 
in insulating AFs vary from system to system.
This indicates that by tuning the laser, we can control and design 
thermomagnetic Hall transport properties in antiferromagnetic magnonic systems.
Those results for the laser-induced magnon motion and Hall transport properties are summarized in Table \ref{tab:table1}.

\section{Estimate for experiments}
\label{sec:Exp}

The development of laser techniques \cite{LaserPhotoExp,LaserPhotoExp2,LaserPhotoExp3}
in quantum optics \cite{GlauberQuantumOptics} is remarkably rapid.
The advanced laser technologies such as optical 
tweezers~\cite{OpticalTweezers}, plasmonics~\cite{LaserPhotoExp,LaserPhotoExp2}, 
near-field~\cite{LaserPhotoExp2}, 
and metamaterials~\cite{Mukai2014APL} enable us to realize the 
various profile of electric and magnetic fields including an ac electric field gradient we considered in this work.

The thermal Hall effect of magnons has been observed in Ref.~\cite{onose}
and experimental evidence for the magnonic spin Nernst effect has been reported in Ref.~\cite{MagnonNernstExp}.
Therefore making use of those measurement techniques,
our theoretical predictions (Table \ref{tab:table1}), i.e., laser-induced magnonic topological phases,
can be experimentally confirmed by measuring Hall currents. 
As seen in Sec. \ref{sec:WF}, the linearly polarized laser can generate helical edge magnon states and induce the magnonic spin Nernst effect,
while the circularly polarized laser can generate chiral edge magnon states and induce the magnonic thermal Hall effect (Table \ref{tab:table1}).
Thereby measuring Hall currents instead of directly observing edge magnon states~\cite{footnote6},
our theoretical predictions can be experimentally confirmed~\cite{footnote7}. 

We estimate the experimental feasibility with 
taking $\mathrm{Cr_{2}O_{3}}$~\cite{CrOexp,CrOexp2} for example 
following Ref.~\cite{KSJD}. 
This material has the spin quantum number $S=3/2$,
$g$-factor $g=2$,
the lattice constant $a=0.5$ nm,
the easy-axis anisotropy ${\cal{K}}= 0.03$ meV, and
the antiferromagnetic nearest-neighbor spin exchange interaction $J = 15$ meV.
The magnon gap arising from the easy-axis spin anisotropy 
amounts to $\Delta = 4$ meV
and the frequency of cyclotron motions becomes $\omega_{\mathrm{c}}= {\cal{O}}(1)$ GHz.
Thereby using a picosecond laser \cite{LaserPico} $\Omega = {\cal{O}}(1)$ THz
and the experimental scheme proposed in Refs.~\cite{KJD,KSJD},
we expect that the laser-induced magnonic topological phases can be 
realized experimentally in the low temperature regime \cite{magnon10mK,footnote8}. 

Lastly, we comment on the heating effect by laser application.
In this paper, we focus on the magnetic insulators with large electronic band gap.
Hence electric excitations by the laser electric field are negligible and consequently, 
the heating through the electron-phonon coupling (e.g., Joule heating) is negligibly small.
Thus we only have to consider the heating problem of the isolated quantum system.
From Ref.~\cite{MagnusMori}, the energy-absorption rate $P$ of the 
isolated quantum system subjected to periodic driving at the frequency $ 
\Omega $ is bounded as 
$P \leq  \hbar   \omega_c ^2  \exp(- \Omega /{\omega_c}) $.
It is exponentially small in the high frequency regime we considered above.
The estimation is given as $  \Omega /{\omega_c} \sim  10^3 $ with the 
parameters $\omega_{\mathrm{c}}= {\cal{O}}(1)$ GHz and $\Omega = {\cal{O}}(1)$ THz.
Therefore, we conclude that heating effects are irrelevant in our systems.





\section{Discussion}
\label{sec:discussion}

Before concluding, we make further discussions on several points of this paper and the future problems.
First, 
the mechanism of our laser-induced magnonic topological phases discussed 
in Sec.~\ref{subsec:system2} is different \cite{DLprivate} from that of the so-called Floquet TIs~\cite{ReviewFloquetTI,FloquetTI}
in the sense that we do not employ Dirac materials \cite{FloquetOkaSan,FloquetDiracFermion,FloquetDiracFermion2} with a relativistic spectrum (i.e., a linear dispersion)
or ac filed-driven resonance across the band gap \cite{FloquetJK,FloquetJK2}.
We remark that Dirac magnons having a linear dispersion 
are available on two-dimensional honeycomb lattices \cite{DiracMagnon}.
In Ref.~\cite{KJD}, we have studied those Dirac magnons in the AC effect.
The correspondence between Dirac magnons in the AC effect and Dirac electrons in the AB effect \cite{FloquetDiracFermion,FloquetDiracFermion2} is straightforward. For example, by simply replacing the Fermi velocity, electric charge, and the AB vector potential~\cite{bohm} with the magnon velocity, $g\mu_{\rm{B}}$, and ${\mathbf{A}}_{\rm{m}}$, respectively,
one can map the equation for Dirac electrons in the AB effect to that for Dirac magnons in the AC effect. 
Compare Eq. (1) of Ref.~\cite{FloquetDiracFermion} with Eq. (D1) of Ref.~\cite{KJD}.
Therefore by applying this mapping to Floquet TIs established in 
Dirac electron systems~\cite{FloquetDiracFermion,FloquetDiracFermion2}, the magnonic analog of the Floquet TIs can be derived theoretically~\cite{WrongFloquetMagnon,SO_FloquetMagnon2,SO_FloquetMagnon3,SO_FloquetMagnon4}.
Moreover, while it is outside the scope of this work since we focus on the magnon dynamics away from the adiabatic limit $ \Omega \gg  \omega_c $, 
the laser-induced resonance across the Landau energy gap of magnons is expected to be generated, 
in the same way as the ac field-driven resonance across the band gap~\cite{FloquetJK,FloquetJK2},
by tuning the laser frequency to the cyclotron frequency of magnons $  
\Omega \approx \omega_c  $, which we leave for the further study~\cite{DLprivate}.

Second, we comment on the effect of the Dzyaloshinskii-Moriya interaction (DMI)~\cite{DM,DM2,DM3}.
When the inversion symmetry is broken, a time-independent DMI indeed can exist and work as a vector potential~\cite{katsura2}
for magnons in the similar way as the AC phase $ {\mathbf{A}}_{\rm{m}}$ 
induced by electric field gradient.
However, a spatially uniform DMI does not give rise to any emergent electromagnetic field that acts as the Lorentz force on magnons and thus should not change the qualitative behavior of magnons obtained in our work. 
We thus conclude that our results, topological phenomena associated with 
the Landau quantization of Floquet magnons, qualitatively remain 
unchanged even in the presence of such DMI. Those topological phenomena 
are stable even with the Rashba-like splitting of the bands provoked by 
the DMI, which retains TRS. 
Since the possible type of DMI strongly depends on the details of the system, 
e.g., the lattice geometry and the magnetic point group, the 
comprehensive study on the effects caused by DMI is beyond the scope of the present 
paper. 
The effect from the interplay of DMI and magnon chirality in AFs has 
been investigated including the optical excitations such as magnon 
photocurrents~\cite{Kishine,Kishine2,AFnonabelian}. 
Those results are helpful for our future study. 

Third, a general treatment of nonequilibrium-driven topological phases in AFs beyond our theoretical framework~\cite{KatsuraZ2} remains an open problem such as disorder effects due to magnetic impurities or the effects of hybridization of spin-up and spin-down magnons due to the symmetry/conservation breaking terms.
While we treat the steady state in terms of the Floquet theory in 
this paper, considering the transient dynamics, thermalization, 
and open systems~\cite{MagnusMori,MagnusMori2} in the laser application 
is an interesting future problem.

Last, applying a laser to magnets is just one of the ways to drive magnets into nonequilibrium. 
We envision that subjecting magnetic systems to various types of nonequilibrium driving, e.g., time-varying thermal environment or charge/heat currents, 
can be versatile means to realize novel topological phases in magnetic systems.

\section{Summary}
\label{sec:summary}

Let us summarize our results. 
Assuming that the total spin along the $z$ axis is conserved, we have established the laser control of magnonic topological phases in the AF by making use of the AC effect on magnons in laser.
Using the Floquet formalism, we have found in the high frequency regime that the linearly polarized laser can generate helical edge magnon states and induce the magnonic spin Nernst effect, while the circularly polarized laser can generate chiral edge magnon states and induce the magnonic thermal Hall effect.
We have thus provided a handle to control and design topological properties of the insulating AF.
Our result for controlling magnonic topological phases by laser provides a new direction for development of magnonics, and will serve as a bridge between two research areas, magnonics and quantum optics.

\acknowledgements
We acknowledge support by 
the Leading Initiative for Excellent Young Researchers, MEXT, Japan (KN),
the startup fund at the University of Missouri (SKK),
the Swiss National Science Foundation under Division II (ST) and 
ImPact project (No. 2015-PM12-05-01) from the Japan Science and Technology Agency (ST).
We (KN) would like to thank K. Usami for useful feedback on the laser experiment
and D. Loss for fruitful discussions about the significance of this work.
KN is grateful to the hospitality of the T. Giamarchi-group (U. Geneva) during his stay
where a part of this work was carried out.

\appendix*

\section{Floquet formalism}
\label{sec:review3}

In this Appendix, for completeness, we provide details of the straightforward calculation of the effective Floquet Hamiltonian.
For a general framework of the Floquet formalism, 
see the review article Ref.~\cite{FloquetReview}.

\subsection{Floquet Hamiltonian and high frequency expansion}
\label{subsec:review3200}

First let us explain the derivation of the Floquet effective model and 
the high frequency expansion. 
This strategy is applicable to general time-periodic systems. 
Assume that the Hamiltonian has a temporal periodicity 
${\cal{H}} (t)  = {\cal{H}} (t+T)$, where $T$ is the period. 
We can perform the Fourier transform on 
the time-dependent Hamiltonian 
\begin{align}
 {\cal{H}} (t)  = \sum_{m \in{\mathbb{Z}}} H_{m} {\rm{e}}^{i m \Omega t},
\label{periodicity2} 
\end{align}
where $\Omega=2\pi/T$. 
Although it is a difficult problem to obtain 
the exact Floquet effective Hamiltonian 
\begin{align}
 \mathcal{H}_{\mathrm{eff}}\equiv\frac{i}{T}
   \ln \mathcal{T}\exp\Big[-i\int_{0}^{T}\mathcal{H}(t)dt\Big],
\nonumber 
\end{align}
where $\mathcal{T}$ is the time-ordering, 
we can calculate it for the high frequency regime 
$\Omega  \gg  \omega_{\rm{c}}$ in the perturbation way 
using the high frequency expansion~\cite{FloquetReview,FloquetOkaSan4}, 
\begin{align}
 {\cal{H}}_{\rm{eff}}= \sum_{n =0}^{\infty } {\cal{H}}_{\rm{eff}}^{(n)}.
 \label{Heff}
\end{align}
Here ${\cal{H}}_{\rm{eff}}^{(n)}$ is the $1/\Omega^n$ order term. 
We give the explicit formula up to ${\cal{O}}(1/\Omega ^2)$,
\begin{align}
{\cal{H}}_{\rm{eff}}^{(0)} =& H_0,   \label{H0}    \\
{\cal{H}}_{\rm{eff}}^{(1)} =&  \frac{1}{\hbar \Omega } \sum_{m=1}^{\infty }\frac{[H_m, H_{-m}]}{m},    \label{H1}    \\
{\cal{H}}_{\rm{eff}}^{(2)} =&  \frac{1}{(\hbar \Omega)^2} \sum_{m \not=0} \Big( \frac{[H_{-m}, [H_0, H_{m}]]}{2m^2}   \nonumber \\
+& \sum_{m' \not=0, m' \not=m} \frac{[H_{-m'}, [H_{m'-m}, H_{m}]]}{3 m m'} \Big).   \label{H2}
\end{align}
The $1/\Omega$ order term Eq.~\eqref{H1} vanishes when the Hamiltonian 
has time-reversal invariance $\mathcal{H} (t)=\mathcal{H} (-t)$ since $[H_m, H_{-m}]=0$.

\subsection{Application to the insulating AF}
\label{subsec:review32}

Next we apply the Floquet theory described 
in Sec.~\ref{subsec:review3200} 
to the insulating AF with the laser application. 

Section \ref{subsec:system1} in the main text:
Each Fourier component $H_{m}$ [Eq.~\eqref{periodicity2}] for the periodic AC vector potential of 
${\mathbf{A}}_{\rm{m}}(t) = ({\cal{E}}/{c})(0, x {\rm{cos}}(\Omega t), 0)$ becomes
\begin{align}
H_0 =& \frac{1}{2m}\Big[p_x^2 + p_y^2 + \frac{1}{2} \Big(\frac{g \mu_{\rm{B}}}{c} \Big)^2   \Big(\frac{\cal{E}}{c} \Big)^2 x^2 \Big]
                    + \Delta,
\nonumber
\\
H_1 =& H_{-1}= \sigma \frac{1}{2m} \frac{g \mu_{\rm{B}}}{c}   \frac{\cal{E}}{c} p_y x,
\nonumber
\\
H_2 =& H_{-2}= \frac{1}{4} \frac{1}{2m}   \Big(\frac{g \mu_{\rm{B}}}{c} \Big)^2 \Big(\frac{\cal{E}}{c} \Big)^2 x^2,
\nonumber
\end{align}
where $ [H_1, H_{-1}] =  [H_2, H_{-2}] =0 $ due to ${\mathbf{A}}_{\rm{m}}(t) ={\mathbf{A}}_{\rm{m}}(-t)$.
Using the high frequency expansion [Eqs. (\ref{H0})-(\ref{H2})], we obtain the effective Floquet Hamiltonian ${\cal{H}}_{\rm{eff}}^{(n)}$ [Eq. (\ref{Heff})] in the main text.

Section \ref{subsec:system2} in the main text:
Each Fourier component $H_{m}$ [Eq.~\eqref{periodicity2}] for the periodic AC vector potential of 
${\mathbf{A}}_{\rm{m}}(t) = ({\cal{E}}/{c})(0, x {\rm{cos}}^2(\Omega t), 0)$ becomes
\begin{align}
H_0 =& \frac{1}{2m}\Big[p_x^2 + p_y^2  + \sigma  \frac{g \mu_{\rm{B}}}{c} \frac{\cal{E}}{c}   p_y x  
+ \frac{3}{8}\Big(\frac{g \mu_{\rm{B}}}{c} \Big)^2   \Big(\frac{\cal{E}}{c} \Big)^2 x^2 \Big]+ \Delta , 
\nonumber\\
H_2 =& H_{-2}= \frac{1}{2m}\Big[ \sigma \frac{1}{2} \frac{g \mu_{\rm{B}}}{c} \frac{\cal{E}}{c}   p_y x   
+ \frac{1}{4} \Big(\frac{g \mu_{\rm{B}}}{c} \Big)^2   \Big(\frac{\cal{E}}{c} \Big)^2 x^2  \Big],
\nonumber\\
H_4 =& H_{-4}= \frac{1}{16}\frac{1}{2m}  \Big(\frac{g \mu_{\rm{B}}}{c} \Big)^2   \Big(\frac{\cal{E}}{c} \Big)^2 x^2,
\nonumber
\end{align}
where $ [H_2, H_{-2}] =[H_4, H_{-4}] = 0 $ due to ${\mathbf{A}}_{\rm{m}}(t) ={\mathbf{A}}_{\rm{m}}(-t)$.
The high frequency expansion for $\Omega \gg  \omega_{\rm{c}}$ 
[Eqs.~\eqref{H0}-\eqref{H2}] 
provides the effective Floquet Hamiltonian up to ${\cal{O}}(1/\Omega ^2)$ as ${\cal{H}}_{\rm{eff}}={\cal{H}}_{\rm{eff}}^{(0)} +{\cal{H}}_{\rm{eff}}^{(2)}$:
\begin{align}
{\cal{H}}_{\rm{eff}}^{(0)} =& \frac{1}{2m}\Big[ p_x^2  +  \Big(p_y  + \sigma  \frac{g \mu_{\rm{B}}}{c} \frac{{\cal{E}}/2}{c} x \Big)^2 \Big]   + \frac{{\cal{F}}_0}{3}  x^2 + \Delta ,
\nonumber\\
{\cal{H}}_{\rm{eff}}^{(2)} =& \frac{t_0}{2m}\Big( p_y + \sigma  \frac{g \mu_{\rm{B}}}{c} \frac{{\cal{E}}}{c} x \Big)^2 
+ {\cal{F}}_2 x^2,
\nonumber
\end{align}
where 
${\cal{F}}_0 =  (3/8) ({g \mu_{\rm{B}}}/{c})^2  ({\cal{E}}/{c} )^2/2m$,
${\cal{F}}_2 =  (1/2\Omega^2) ({1}/{16})^2 ({1}/{2m})^3 ({g \mu_{\rm{B}}}/{c})^4  ({\cal{E}}/{c})^4$, and 
$t_0 =  (1/32)({\omega _c}/{\Omega})^2  \propto  1/{\Omega^2}$.
It is rewritten as 
\begin{align}
 {\cal{H}}_{\rm{eff}}=&\frac{1}{2m}\Big[ p_x^2 + (1+t_0)\Big(p_y + \sigma \frac{g \mu_{\rm{B}}}{c} \frac{{\cal{E}}_{\rm{eff}}}{c}x \Big)^2 \Big] 
\nonumber\\
&+ {\cal{F}} x^2+ \Delta ,
 \label{Heff2total_Appendix}
\end{align}
where 
${\cal{F}} =  {\cal{F}}_0 + {\cal{F}}_2 + {\cal{F}}_t - {\cal{F}}_3$,
${\cal{F}}_t =  ({g \mu_{\rm{B}}}/{c})^2 ({\cal{E}}/{c})^2 t_0/2m$,
${\cal{F}}_3 =  [{(1+2 t_0)^2}/({1+t_0})] ({g \mu_{\rm{B}}}/{c})^2  ({\cal{E}}/{c})^2/8 m$.
The effective electric field gradient in Eq.~\eqref{Heff2total_Appendix} is given by
$  {\cal{E}}_{\rm{eff}} = [(1+2 t_0)/(1+t_0)] {\cal{E}}/2 $.

Section \ref{subsec:system3} in the main text:
Each Fourier component $H_{m}$ [Eq.~\eqref{periodicity2}] for the periodic AC vector potential of 
${\mathbf{A}}_{\rm{m}}(t) = ({\cal{E}}/{c})(\eta  x {\rm{sin}}(\Omega t), x {\rm{cos}}(\Omega t), 0)$ becomes
\begin{align}
H_0 =& \frac{1}{2m}\Big[p_x^2 + p_y^2 +  \Big(\frac{g \mu_{\rm{B}}}{c} \Big)^2   \Big(\frac{\cal{E}}{c} \Big)^2 x^2 \Big]+ \Delta,
\nonumber\\
H_1 =& - \frac{\hbar }{4 m}  \sigma  \frac{g \mu_{\rm{B}}}{c} \frac{\cal{E}}{c} \eta  
+ \frac{1}{2m}\sigma  \frac{g \mu_{\rm{B}}}{c} \frac{\cal{E}}{c}(-i \eta  x p_x + x p_y),
\nonumber\\
H_{-1} =&   \frac{\hbar }{4 m}  \sigma  \frac{g \mu_{\rm{B}}}{c} \frac{\cal{E}}{c} \eta  
+ \frac{1}{2m}\sigma  \frac{g \mu_{\rm{B}}}{c} \frac{\cal{E}}{c}(i \eta  x p_x + x p_y),
\nonumber
\end{align}
where $ [H_1, H_{-1}] \neq 0 $ due to ${\mathbf{A}}_{\rm{m}}(t) \neq{\mathbf{A}}_{\rm{m}}(-t)$.
Using the high frequency expansion [Eqs.~\eqref{H0}-\eqref{H2}], 
we obtain ${\cal{H}}_{\rm{eff}}^{(0)} $ and ${\cal{H}}_{\rm{eff}}^{(1)}$ in the main text,
and the effective Floquet Hamiltonian ${\cal{H}}_{\rm{eff}}={\cal{H}}_{\rm{eff}}^{(0)} +{\cal{H}}_{\rm{eff}}^{(1)}$ is rewritten as 
\begin{align}
 {\cal{H}}_{\rm{eff}}=& \frac{1}{2m}\Big[ p_x^2 + \Big(p_y - \eta  \frac{g \mu_{\rm{B}}}{c} \frac{{\cal{E}}_{\rm{eff}}}{c}x \Big)^2 \Big]  
+ \Delta 
 \nonumber   \\
+&  \frac{1}{2m}\Big[1- \Big( \frac{\omega_c}{2 \Omega }\Big)^2 \Big]  \Big(\frac{g \mu_{\rm{B}}}{c} \Big)^2   \Big(\frac{\cal{E}}{c} \Big)^2 x^2.
\nonumber
\end{align}

\bibliography{PumpingRef}

\begin{thebibliography}{116}
\bibitem[{foo({\natexlab{b}})}]{footnote2}
\bibinfo{note}{The force acting on magnons in the dc electric field
  [Eq.~\eqref{HamiltonianAC}] is invariant under a gauge transformation
  \cite{KSJD} and the gauge invariance is specific to only magnons, i.e.
  electrically neutral particles.}
\bibitem[{foo({\natexlab{c}})}]{footnote4}
\bibinfo{note}{The magnon gap due to the easy-axis spin anisotropy amounts to
  $\Delta = 4$ meV for~\cite{CrOexp,CrOexp2} $\mathrm{Cr_{2}O_{3}}$, see
  Ref.~[\onlinecite{KSJD}] for details of experiment parameter values.Compared
  with it, the magnetic energy due to the resulting ac magnetic field from the
  applied laser (i.e., ac electric field),e.g., picosecond
  laser~\cite{LaserPico} $\Omega = \mathcal{O}(1)$ THz, is negligibly small.
  Furthermore, the magnetic energy can be substantially softened by allowing
  for a periodic extension of linear electric field gradients~\cite{KJD,KSJD}.}
\bibitem[{foo({\natexlab{f}})}]{footnote6}
\bibinfo{note}{We expect that neutron scattering
  measurements~\cite{ExpPSI,TopoMagBandKagome} will be one of the promising
  strategies for the observation of magnon edge states.}
\bibitem[{foo({\natexlab{g}})}]{footnote7}
\bibinfo{note}{The AC effect in magnetic systems has also been experimentally
  confirmed in Ref.~[\onlinecite{ACspinwave}].}
\bibitem[{foo({\natexlab{h}})}]{footnote8}
\bibinfo{note}{Note that even a femtosecond laser pulse is now
  available~\cite{Laser}.}
\end{thebibliography}

\end{document}